\documentclass[12pt,preprint]{aastex}

\received{}
\accepted{}

\slugcomment{Submitted to the Astrophysical Journal}

\shorttitle{Jet of M87}
\shortauthors{Wilson \& Yang}

\begin{document}

\title{Chandra X-ray Imaging and Spectroscopy of the M87 Jet and Nucleus}

\author{A. S. Wilson\altaffilmark{1} and Y. Yang}
\affil{Astronomy Department, University of Maryland, College Park,
MD 20742; wilson@astro.umd.edu, yyang@astro.umd.edu}


\altaffiltext{1}{Adjunct Astronomer, Space Telescope Science Institute,
3700 San Martin Drive,
Baltimore, MD 21218; awilson@stsci.edu}


\begin{abstract}

We report X-ray imaging - spectroscopy of the jet of M87 at sub arc second
resolution with the Chandra X-ray Observatory. The galaxy nucleus and all the
knots seen at radio and optical wavelengths, as far from the nucleus as knot
C, are detected in the X-ray observations. There is a strong trend for the
ratio of X-ray to radio, or optical, flux to decline with increasing
distance from the nucleus. At least three knots are displaced from their
radio/optical counterparts, being tens of pc closer to the nucleus at X-ray
than at radio or optical wavelengths. The X-ray spectra of the nucleus and 
knots are well described by power laws absorbed by cold gas, with only the
unresolved nucleus exhibiting intrinsic absorption. In view of the similar
spectra of the nucleus and jet knots, and the high X-ray flux of the knots
closest to the nucleus, we suggest that the X-ray emission coincident with
the nucleus may actually originate from the pc -- or sub-pc -- scale jet
rather than the accretion disk.

Arguments are given that the X-ray emission process is unlikely to be inverse
Compton scattering. Instead, we favor synchrotron radiation. 
Plotted as $\nu$S$_{\rm \nu}$, the spectra of the knots generally peak in or 
just above the optical - near infrared band. However, the overall spectra of at
least three knots cannot be described by simple models in which the spectral
index monotonically increases with frequency, as would result from synchrotron
losses or a high energy cut-off to the injected electron spectrum. Instead,
these spectra must turn down just above the optical band and then flatten 
in the X-ray band. In the context of a synchrotron model, this result
suggests that either the X-ray emitting electrons/positrons in these knots
represent a separate ``population'' from those that emit the radio and optical
radiation or the magnetic field is highly inhomogeneous. If the former
interpretation is correct,
our results provide further support
for the notion that radio galaxies produce a hard ($\gamma$ $\simeq$ 2 -- 2.5,
N(E) $\propto$ E$^{-\gamma}$) spectrum of high energy (E/m$_{e}$c$^{2}$)
$\sim$ 10$^{7-8}$) electrons and possibly positrons.

\end{abstract}




%

\newpage
\section{INTRODUCTION}

The jet in M87 is arguably the most famous of its kind, primarily because of
its proximity and strong optical synchrotron radiation. It has been studied in
great detail at radio (see Biretta 1999 for a review) and optical (e.g.
Biretta, Sparks \& Macchetto 1999; Perlman et al. 2001) wavelengths. Radio
observations of the jet show that it is polar down to 0.01 pc (Junor \&
Biretta 1995), which is only 60 times the gravitational radius of the
black hole (mass $\simeq$ 3 $\times$ 10$^{9}$ M$_{\odot}$, e.g. Macchetto
et al. 1997). 
HST observations have revealed numerous features in the inner $\simeq$
400 pc with speeds in the range 4c to 6c, confirming that the bulk flow of the
jet is
relativistic (Biretta, Sparks \& Macchetto 1999).

At X-ray wavelengths, the nucleus and knot A have been detected (Schreier,
Gorenstein \& Feigelson 1982; Biretta, Stern \& Harris 1991; Neumann et al.
1997; Harris, Biretta \& Junor 1997; B\"ohringer et al. 2001). The limited
resolution of prior
X-ray telescopes has precluded a study with comparable resolution to optical
telescopes. With the advent of the Chandra X-ray Observatory, it is now
possible to obtain imaging - spectroscopy of the jet with sub arc second
resolution.

In Section 2, we describe the Chandra observations and the data reduction.
Section 3 presents the results, discussing the morphology of the jet and the
spectra of the nucleus and individual jet knots. Section 4 is devoted to the
radiation mechanism of the X-rays, while Section 5 summarises
our conclusions.
We assume a distance of 16 Mpc (Tonry 1991; Whitmore et al. 1995), so
1$^{\prime\prime}$ = 78 pc.
The Galactic column density in the direction of M87 is N$_{\rm H}$ (Gal) =
2.5 $\times$ 10$^{20}$ cm$^{-2}$ (Stark et al. 1992).

\section{OBSERVATIONS AND DATA REDUCTION}

Because the nucleus and knot A of the jet are known to be strong X-ray emitters,
we were concerned that use of the default 3.2s frame time for ACIS would
result in pile-up in these and possibly other regions. We therefore obtained
a short ($<$ 1 ks) exploratory observation of M87 on April 20 2000 (obsid 351)
with the nucleus at the aim point on ACIS-S (chip S3). This observation was made
in so called ``alternating mode'', with the frame time alternating between
0.1 and 0.4s. This observation showed the peak, level 2 count rate to be
$\simeq$ 0.10 cts s$^{-1}$ pixel$^{-1}$ for both frame times, implying that 
observations with a 3.2s frame time would suffer significantly from pile-up,
but that observations with a 0.4s frame time would be free of it.

The science observations were, therefore, taken in two parts. First, an 
observation with a 3.2s frame time was obtained on July 29 2000 (obsid 352).
CCDs I2, I3,
S1, S2, S3 and S4 were on and the good exposure time (LIVTIME) was 37.6 ks.
This observation was designed for study of the cluster and the fainter parts of
the jet.
Second, an observation with 0.4s frame time was taken on July 30 2000
(obsid 1808). Only S3 was used, and the exposed region comprised a band
127 rows (62$^{\prime\prime}$) wide with length the whole chip. The width of
the band suffices to include the whole jet. The good exposure time was 
12.8 ks. 
The peak count rate at the nucleus in this exposure is 
0.07 cts s$^{-1}$ pixel$^{-1}$, which is different to that obtained in the
exploratory exposure, suggesting 
variability of
the nucleus between April 20 and July 30 2000 (the corresponding total count
rates from the nucleus are 0.55 and 0.39 cts s$^{-1}$ for these two dates,
respectively). This count rate confirms that 
the nucleus is not piled-up in a 0.4s frame time. From this second exposure,
we have obtained spectra of
the brighter regions in the field, specifically the nucleus and the brighter
parts of the jet.

We followed the
procedures recommended in the ``science
threads'', including filtering of data for times of 
high background and aspect errors,
of which there were very few,
and construction of response matrix functions and ancillary response files
at the locations of the nucleus and knots of the jet.
In order to improve the sampling of the image, we interpolated to a pixel
size of 0\farcs1 and then smoothed the data with a Gaussian of FWHM =
0\farcs5.
Processing of the data initially used ciao 1.1.5, but on March 19 2001 
reprocessed data were received, which must be processed with ciao 2. We
made a careful comparison of the results of processing the old data with
ciao 1.1.5 and the new data with ciao 2 and found no significant
differences. 

While this paper was being written (July 2001), preliminary versions of new
response functions (``fefs'') became available. We reprocessed the spectra of
knot A with these new functions and found that all parameters of the absorbed 
power law model spectra (N$_{\rm H}$, $\Gamma$ and K, see Table 2) were within
the 90\% confidence error ranges of the parameters obtained with the old
response functions. For uniformity, we give only parameters obtained with the
old response functions, which are quite adequate for dealing with continuum
spectra, as found in the M87 jet.

\section{RESULTS} 

\subsection {Astrometry and Morphology}

A grey scale image and a contour map of the jet (from the 3.2s frame time
observation) are shown in Fig. 1, while Fig. 2 compares radio, optical and
X-ray images with similar FWHM resolutions - 0$\farcs$4, 0$\farcs$7
and $\simeq$ 0$\farcs$7, respectively. 
The position
of the X-ray nucleus (in J2000.0 coordinates) is RA$_{\rm x}$ = 
12$^{h}$ 30$^{m}$ 49\fs40, Dec$_{\rm x}$ = 
12$^{\circ}$ 23$\arcmin$ 27$\farcs$8. 
This may
be compared with two VLA radio positions -
RA$_{\rm r1}$ = 12$^{h}$ 30$^{m}$ 49\fs423, Dec$_{\rm r1}$ =
12$^{\circ}$ 23$\arcmin$ 28\farcs11 from an archival 15 GHz map made by Carole 
Mundell
and RA$_{\rm r2}$ = 12$^{h}$ 30$^{m}$ 49\fs423, Dec$_{\rm r2}$ =
12$^{\circ}$ 23$\arcmin$ 28\farcs02 from a 5 GHz map provided by John
Biretta.
The differences are RA$_{\rm x}$ -- RA$_{\rm r1}$ = --0\farcs3,
Dec$_{\rm x}$ -- Dec$_{\rm r1}$ = --0\farcs3, and
RA$_{\rm x}$ -- RA$_{\rm r2}$ = --0\farcs3,
Dec$_{\rm x}$ -- Dec$_{\rm r2}$ = 
--0\farcs2, which represent excellent agreement. No attempt has been made to 
remove the X-ray emission associated with
knot HST-1, which has probably ``pulled''
the nuclear peak slightly westward; correction for this effect may improve
the agreement in right ascension.

A casual examination of Fig. 2 shows the
X-ray image to be similar to those in the other wavebands, with the nucleus,
and jet knots HST-1, D, E, F, I, A, B and C (listed in order of increasing
distance from the nucleus) all visible (we follow the
traditional nomenclature, see e.g. Perlman et al. 2001). However, it is clear
that the knots closest to the nucleus - HST-1 and D - are much brighter
relative to the other knots than at radio and optical wavelengths. We discuss
the spectral differences between the knots quantitatively in Section 3.2.

The centroid position of each knot has been measured using the iraf task 
`center'. We have also determined the position of the peak of emission through
interpolation using the 3 $\times$ 3 pixels centered on the brightest one.
These two positions do not always agree because of internal structure and
asymmetries in the brightness distributions of individual knots. Table 1
lists the distance of each knot from the nucleus in each waveband. When two 
numbers are listed separated by a dash, they represent the distances of the
centroid and the peak from the nucleus, giving an approximate idea of the
range resulting from internal structure. It is clear
from Table 1 and the intensity profiles along the jet plotted in Fig. 3 
that the X-ray emission
of knots D and F is closer to the nucleus than at radio and optical wavelengths
at these spatial resolutions. The X-ray emission of knots E and B may also be
closer to the nucleus than their radio and optical counterparts, but here the
effect is marginal.
Comparison with the much higher resolution HST
optical observations (Perlman et al. 2001) shows that the X-ray emission of 
knot D peaks close to the optical emission of sub-knot D-East, while the
X-ray emission of knot F is located near the upstream edge of the optical knot.
Measurement of
the position of knot A is complicated by the asymmetric structure found in
higher resolution (0$\farcs$1 - 0$\farcs$2) maps than shown here at both radio
(e.g. Owen, Hardee \& Cornwell 1989) and optical (Perlman et al. 2001)
wavelengths. However, the present observations show that the distance from the 
nucleus to knot A is the same at all three wavelengths to within $\pm$ 
0$\farcs$1. This conclusion differs from that of Neumann et al. (1997) who
found, based on ROSAT observations, that the X-ray maximum of knot A is 
situated 0$\farcs$4 nearer to the nucleus than the radio maximum. 
Neumann et al.'s (1997) finding undoubtedly results from the inclusion of the
X-ray bright knots HST-1 and D in their nuclear source given the 
5$^{\prime\prime}$ resolution of the ROSAT HRI. If the Chandra image is 
smoothed to this resolution, the separation of the ``nuclear'' and knot
A peaks becomes 11$\farcs$6, in reasonable agreement with Neumann et al.'s
(1997) result.

Fig. 3 shows profiles of the radio, optical and X-ray brightnesses along the 
jet.
The brightnesses have been normalised to be the same at knot A. In this
representation, the X-ray emission lies above the radio and optical closer to 
the nucleus than knots A and F (i.e. for knots HST-1, D and E) but below the 
radio and optical beyond knot A (i.e. for knots B and C). The X-ray emission 
of the jet thus declines more rapidly with increasing distance from the nucleus
than does the radio or optical emission.

\subsection {Spectra}

X-ray spectra have been obtained for the nucleus and all knots out to knot C.
Background spectra were taken for each knot from two regions just outside the
jet and along a line perpendicular to it passing through the knot in question.
The spectra obtained are insensitive to the exact choice of background since
the jet is very bright compared to its surroundings.
The spectra of all knots were modelled with a power law
absorbed by cold matter. The resulting equivalent hydrogen column density,
photon index and normalisation of the spectra are given in Table 2. 
Our photon indices for the nucleus and knot A are in excellent agreement
with those obtained by XMM (B\"ohringer et al. 2001), though the latter
data average over a larger area. Knot
HST-1 is significantly contaminated by nuclear emission so we do not list
the parameters of its spectrum. However, its spectrum appears to be well
described by a power law with index $\Gamma$ = 2.24, similar to the rest of the
jet. Deconvolution (deferred to a later paper) will be necessary to obtain a 
more reliable measure of its spectral parameters.

Of the jet knots, only knot F shows evidence of an absorbing column in excess
of the Galactic column of N$_{\rm H}$ (Gal) =
2.5 $\times$ 10$^{20}$ cm$^{-2}$. This possible excess should be treated with
caution in view of uncertainties in the instrumental response below 0.4 keV.
On the other hand, the nucleus shows clear excess absorption of $\simeq$
(3 -- 5) $\times$ 10$^{20}$ cm$^{-2}$ over the Galactic column (Table 2).
Comparison of the model absorbed and unabsorbed spectra shows that this
absorption is too large to be an instrumental error. 

The photon indices of the nucleus and all knots are quite similar (Table 2),
ranging between best estimate values of $\Gamma$ = 2.04 and 2.90 (flux
spectral indices $\alpha$ between 1.04 and 1.90). The similarity of the nuclear
spectrum to that of the jet and the strong increase of the brightness of the
X-ray jet towards the nucleus (Fig. 3)
suggests that the ``nuclear'' emission may actually
originate from the pc -- or sub-pc -- scale jet.
Various VLBI observations have imaged
the jet down to milli arc second scales (e.g. Junor \& Biretta 1995), and it
is reasonable that there should be X-ray emission from the jet unresolved
by Chandra.

Fig. 4 shows radio through X-ray spectra for eight knots plus the nucleus.
Plotted as $\nu$S$_{\nu}$, the spectra are seen to turn over in or
just above the optical-near infrared band, a phenomenon sometimes attributed to
synchrotron losses
(e.g. Perlman et al. 2001 and references therein). 
These broad band spectra are discussed later in the context
of the X-ray emission process (Section 4).

Fig. 5 is a plot of the optical to X-ray ($\alpha_{\rm ox}$), radio to X-ray
($\alpha_{\rm rx}$) and X-ray ($\alpha_{\rm x}$) spectral indices of the knots 
as a function of distance from the nucleus. As can be seen, both
$\alpha_{\rm ox}$ and $\alpha_{\rm rx}$ increase systematically with distance 
from the nucleus (the errors in these spectral indices are smaller than the
symbol sizes). This is another way of saying that the X-ray emission declines
more rapidly with increasing distance from the nucleus than does the radio
or optical emission (Section 3.1 and Fig. 3).

\section {The Nature of the X-ray Emission}
\subsection {Inverse Compton Scattering}

A plausible process for X-ray emission from radio jets is inverse Compton
scattering. The photons being upscattered can either be the synchrotron 
radiation (synchrotron self-Compton [SSC] process) or some external source
(external Compton [EC]), such as the galaxy starlight, radiation from
the active nucleus or the microwave background. The possibility of bulk 
relativistic motion by the jet needs to be considered (see e.g. Begelman,
Blandford \& Rees 1984 for a review). We consider the cases with and without
relativistic motion in the next two subsections.

\subsubsection{No bulk relativistic motion}

We have performed numerical calculations of spectra in spherical geometries 
using the computer code of Band \& Grindlay (1985, 1986), which was kindly
provided by D. E. Harris. Equipartition fields for the knots are given by
Owen, Hardee \& Cornwell (1989) and are in the range (1.6 -- 5.1) $\times$
10$^{-4}$ gauss. We have calculated inverse Compton models for knot B, which
should be representative of the other knots. Assuming the radio to optical
spectrum given in Fig. 4 and low and high frequency cut-offs to the synchrotron 
radiation at 10$^{7}$ and 10$^{15}$ Hz (see Fig. 4 and Meisenheimer et al. 
1996), we find the internal radiation density from synchrotron radiation to be
$\simeq$ 8 $\times$ 10$^{-10}$ erg cm$^{-3}$, which exceeds other known sources
of radiation (an exception might be a narrow beam of radiation
from the nucleus and aligned with the jet, but we ignore this possibility
here). The radiation density of the synchrotron radiation is, however, an order
of magnitude below that of the equipartition magnetic field
(B$_{\rm eq}$ $\simeq$ 3 $\times$ 10$^{-4}$) plus relativistic particles,
indicating the electrons/positrons lose energy mostly to synchrotron radiation.
The calculation shows that the SSC radiation is $\simeq$ 600 times smaller than
observed at 1 keV for an equipartition field. The emissivity
$\epsilon_{\nu}^{\rm SSC}$ $\propto$ N$_{\rm e0}$ 
$\propto$ B$^{-(\gamma + 1)/2}$ (where $\gamma$
is the exponent in the electron
energy spectrum N(E) = N$_{\rm e0}$E$^{-\gamma}$)
for a given synchrotron emissivity. Thus the field strength would have to
be a factor $\simeq$ 70 times weaker than equipartition (given that
$\alpha$ = 0.48 (S$_{\nu}$ $\propto$ $\nu^{-\alpha}$) for knot B) for 
$\epsilon_{\nu}^{\rm SSC}$ to agree with the emissivity observed at 1 keV.

For inverse Compton scattering off the microwave background, the predicted 
intensity at 1 keV is 10$^{5-6}$ lower than observed for an equipartition
field. These results render inverse Compton models extremely unlikely in the 
absence of bulk relativistic motion.

\subsubsection{Bulk relativistic motion}

The above considerations must be modified if the material comprising the
jet knots is moving relativistically. There is now clear evidence for such
relativistic motions in the first 6$^{\prime\prime}$ of the jet based on HST
observations of apparent speeds of features in the range 4c -- 6c
(Biretta, Sparks \& Macchetto 1999).

As usual, we use the Doppler
factor $\delta$ (where $\delta = [L(1\ - \beta\ \cos\theta)]^{-1}$,
$L$ is the Lorentz factor of the bulk flow, $\beta$ is the
bulk velocity in units of the speed of light, and $\theta$ is
the angle between the velocity vector and the line of sight).
The expression for the ratio of SSC X-ray spectral flux to synchrotron
radio spectral flux is often written in terms of
parameters defining the synchrotron
self absorption (SSA) turnover, such as $\nu_{\rm m}$ (frequency of SSA
turnover) and S$_{\rm m}$ (flux density at $\nu_{\rm m}$, extrapolated from the
optically thin spectrum). In this case, S$_{\nu, \rm SSC}$
$\propto$ $\delta^{-2(\alpha + 2)}$ (Burbidge, Jones \& O'Dell 1974;
Marscher 1983). This parameterisation results from the historical tendency
for the SSC model to be applied  to compact radio sources with detected
X-ray emission, for which $\nu_{\rm m}$ and S$_{\rm m}$ are directly
measured (e.g. Marscher et al. 1979). However, in the case of extended jets and
hot spots, SSA almost always occurs at too low a frequency to be observed,
and it is more intuitive to express S$_{\nu, \rm SSC}$ in terms of N$_{\rm e0}$
and B. In this case, the dependence of SSC flux density
on $\delta$ for isotropic radiation in the
blob's rest frame is
S$_{\nu, \rm SSC}$ $\propto$ $\delta^{3 + \alpha}$. 
This behavior is identical to the dependence for
synchrotron radiation (also assumed isotropic in the blob's rest frame),
S$_{\nu, \rm synch}$ $\propto$ $\delta^{3 + \alpha}$,
so the ratio S$_{\nu, \rm SSC}$/S$_{\nu, \rm synch}$ $\propto$ N$_{\rm e0}$
(N$_{\rm e0}$ is measured in the blob's frame)
and is independent of $\delta$ (Dermer, Sturner \&
Schlickeiser 1997). We thus see that S$_{\nu, \rm SSC}$/S$_{\nu, \rm synch}$
is unaffected by bulk relativistic motion given the assumption that the
radiation is isotropic in the blob's rest frame,
and that the only way to increase
the SSC flux density for a given radio flux density is to increase
N$_{\rm e0}$ (or, equivalently, to decrease B [Sect 4.1.1]).
Recalling S$_{\nu, \rm synch}$ $\propto$ $\delta^{3 + \alpha}$ B$^{1 + \alpha}$
N$_{\rm e0}$ V$_{\rm b}$ (where V$_{\rm b}$ is the volume of the blob) and
assuming S$_{\nu, \rm synch}$ is a fixed observable, then B must decrease if
either $\delta$ or N$_{\rm e0}$ increase. For fixed S$_{\nu, \rm synch}$ and
S$_{\nu, \rm synch}$/S$_{\nu, \rm SSC}$,
B $\propto$ $\delta^{-(3 + \alpha)/(1 + \alpha)}$ $\propto$ $\delta^{-2.1}$ for
$\alpha$ = 0.75, in agreement with Tavecchio et al. (2000, their Fig. 1).

The situation is different for inverse Compton scattering of the microwave
background. Here the radiation is isotropic in the comoving Hubble frame and
anisotropic in the frame of the blob. Now the principal dependence of the
external Compton scattered flux density is
S$_{\nu, \rm EC}$ $\propto$ $\delta^{4 + 2\alpha}$ (Dermer 1995;
Dermer, Sturner \& Schlickeiser 1997). Thus
S$_{\nu, \rm EC}$/S$_{\nu, \rm synch}$ $\propto$ $\delta^{1 + \alpha}$ so the
EC X-ray flux is boosted relative to the synchrotron radio flux if
$\delta$ $>$ 1. 
Harris \& Krawczynski (2002) have derived the beaming parameters required
in this model, assuming equipartition magnetic fields. Their values of
$\delta$ range from 11 to 40 and $\theta$ from 1\fdg2 to 4\fdg7. This
combination is different from those required to explain the observed 
superluminal motions (up to 6c - see Biretta et al. 1999), as Harris \& 
Krawczynski note. Values of $\theta$ as small as 1$^{\circ}$ - 5$^{\circ}$
are difficult to reconcile with the clean separation on the sky of the inner
radio lobes of M87.

Thus we conclude that inverse Compton models fail unless the magnetic field
is substantially weaker than equipartition. In addition, inverse
Compton radiation should have the same spectral index as power-law synchrotron
radiation (away from the ``end points''). However, Table 2 and Fig. 4 show that 
the X-ray spectrum is always steeper than the radio spectrum. 
A way out of this problem is to postulate the existence of a steep spectrum, low
energy population of relativistic electrons. These electrons would radiate
synchrotron radiation at lower frequencies than those at which the jet has been
observed and would inverse Compton scatter photons to X-ray energies. 
Such models are {\it ad hoc}
and also require that the low energy electron population is in an
extremely weak magnetic field (see e.g. model 3, Table 4 and Fig. 9 of Wilson, 
Young \& Shopbell [2001], which attempts to explain the properties of
the western hot spot of Pictor A in terms of such a model). 
While it is impossible to rule out such models absolutely (the relevant
relativistic electrons could be in a region of weak or no magnetic field
and thus radiate little or 
no synchrotron radiation), we consider all inverse Compton 
scattering models for the
X-ray emission of the M87 jet to be implausible.

\subsection {Synchrotron Radiation}

The X-ray emission of the jet is almost certainly synchrotron radiation, as
preferred by several earlier workers (e.g. Biretta et al. 1991; Neumann et al. 
1997). This hypothesis is favored by the steep X-ray 
spectra of the knots (Fig. 4), 
which
presumably result, at least in part, from synchrotron losses given that the
electrons have
E/m$_{\rm e}$c$^{2}$ $\sim$ 10$^{7-8}$ 
and half lives of order years. The optical spectra
are also steeper than the radio, a result which has also been 
ascribed to synchrotron losses
(e.g. Perlman et al. 2001).
Our finding that both $\alpha_{\rm ox}$ and $\alpha_{\rm rx}$ increase
systematically
with increasing distance from the nucleus suggests a trend of either
increasing synchrotron losses or decreasing high energy particle acceleration
at larger nuclear distances. The latter effect might be related to 
deceleration of the jet.

Perlman et al. (2001) have shown that simple models involving synchrotron
losses can match the optical spectra of individual knots. They considered a 
variety of possible situations and were able to exclude a model for knot A
in which the
electrons are continuously injected (CI) with the index of the electron
energy spectrum being the same over the whole range of radio to X-ray emitting
energies. In the CI model, both low and high frequency regimes are power
laws with a difference in spectral index $\Delta\alpha$ = 0.5 (Ginzburg 1957;
Kardashev 1962). Such a model that correctly accounts for the radio -- optical 
spectrum of knot A overpredicts its X-ray flux. Perlman et al. (2001) showed
that other models (e.g. Kardashev 1962; Jaffe \& Perola 1973), in which there
is no continuous injection of relativistic electrons at high energies, are able
to account for the radio -- optical spectra of several knots, with reasonable 
concurrence with the X-ray flux point then known for knot A. These authors also
suggested that the weak X-ray flux might result if, for some reason, the
corresponding relativistic electrons fill volumes much smaller than the optical
emitting regions.

In general, the effect of a high energy cut-off or synchrotron/inverse
Compton losses on the relativistic electron
energy spectrum in a magnetic field of uniform strength
is a synchrotron spectrum
exhibiting a monotonically increasing slope (spectral index)
with increasing frequency.
Examination of the multi-band spectra (Fig. 4) shows that the radio through
X-ray spectra of
some knots (e.g. E, F, I) may be qualitatively consistent with
this expectation.
However, in other cases (knots A, B and probably D), the X-ray spectrum cannot
be reproduced by extending the radio through optical fluxes with a spectrum
having a monotonically increasing spectral index with increasing energy. In 
these cases, the spectrum must turn down above the optical, and then 
flatten out to match the X-ray spectrum. This situation is identical to that
found for the western hot spot of Pictor A (Fig. 8 of Wilson, Young \&
Shopbell 2001). 

There are three general ways in which such spectra may arise, as we now 
discuss:

\noindent
(i) If the magnetic field does not depart greatly from uniform strength within
any given knot, the X-rays must come from a different ``population'' of
relativistic electrons from those responsible for the radio and optical 
emissions. In this case, the spectra turn down sharply above the optical band
due to synchrotron losses or a high energy cutoff in the injection spectrum,
with a separate ``population'' dominating the X-ray emission.

\noindent
(ii) The magnetic field departs greatly from uniformity.

\noindent
(iii) The X-ray and/or optical emission is highly variable, so that the X-rays
have declined or the optical increased between the dates the optical
(Feb or April 1998 for Perlman et al.'s [2001] HST data, the triangles in
Fig. 4) and X-ray (July 30 2000) observations were made. 

We discuss these three
possibilities in turn.

In possibility (i), the index of the electron spectrum at injection 
$\gamma_{\rm x}$ = 2$\alpha_{\rm x}$ for the X-ray-emitting electrons/positrons
(assuming an equilibrium between injection and synchrotron losses).
Thus $\gamma_{\rm x}$ ranges between 2.08 and 3.80.
The
latter value of $\gamma_{\rm x}$ (for knot F) has large errors and most
values of $\gamma_{\rm x}$ are $\simeq$ 2.5 -- 2.6 (Table 2). These numbers
agree well with the latest expectations from both diffusive
shock acceleration theory,
incorporating anomalous transport, 
which gives $\gamma$ $\simeq$ 2.2 - 2.5 (Kirk \& Dendy 2001),
and acceleration by relativistic shocks, which gives $\gamma$ = 2.23 in the
ultrarelativistic limit (Kirk 2002).
As well as being directly accelerated by shocks, high energy relativistic
electrons may result from a ``proton-induced cascade'' initiated by
photopion production; this process gives a synchrotron X-ray spectrum which
is distinct from extrapolations of the synchrotron 
radio - optical spectrum (e.g. Mannheim,
Kr\"ulls \& Biermann 1991). The notion that the X-ray emission results from a
separate, high energy component of relativistic electrons/positrons 
is consistent with the different morphologies and locations
of some X-ray knots compared to the corresponding radio and optical
knots (Section 3.1). In an earlier paper (Wilson, Young \& Shopbell 2001),
we noted, based on a very small sample, that {\it hot spots} which are
overluminous in X-rays compared
with the predictions of SSC models appear to be in environments with low
intergalactic gas density. One speculative interpretation is that the
resulting high outward velocity of the hot spots is conducive to acceleration
of the
E/m$_{\rm e}$c$^{2}$ $\sim$ 10$^{7-8}$ 
particles needed for X-ray synchrotron
emission. Such a scenario could also apply to the {\it jet} of M87 since
superluminal proper motions have been detected out to 6$^{\prime\prime}$
from the nucleus, strongly suggesting the bulk motions of the jet are
relativistic (Biretta et al. 1999).

Turning to possibility (ii), there is some literature on the properties of
synchrotron radiation in inhomogeneous magnetic fields (e.g. Cavallo, Horstman
\& Muracchini 1980). An extreme proposal is Burn's (1973) model of the Crab
Nebula. Burn proposed that there is a large scale field, superimposed on which
are regions of higher magnetic field strength, perhaps filamentary in 
morphology (the ``wisps'' would be examples) and produced by hydromagnetic
disturbances, the energy source for which is the pulsar. The model does not
require the usually-supposed continuous injection of high energy 
relativistic electrons. Instead, the radiation at frequencies higher than 
optical originates from electrons that are temporarily caught in regions of 
enhanced magnetic field, but which spend most of their time in lower fields
when they radiate at radio and optical frequencies. Only a small fraction of the
nebula has these high field regions, which must have strengths
more than 100 times the 
undisturbed field in order to account for the extended X-ray emission of the
nebula. The spectrum of the high frequency emission then reflects the
spectrum of the magnetic field strength rather than that of the
relativistic electrons. Since the magnetic field strength in the Crab and
the M87 knots are similar (several $\times$ 10$^{-4}$ gauss), the
qualitative character of the model would be similar in M87, with the 
hydromagnetic compressions being driven by the jet. In particular, the observed 
flat X-ray spectra of knots A, B and D would be related to the spectrum of the
magnetic field strength. State-of-the-art, three dimensional MHD simulations
of radio galaxies (e.g. Tregillis, Jones \& Ryu 2001) can, in principle,
address the crucial question of whether the very large enhancements of magnetic
field strength required in Burn's model are likely to occur in nature.

For possibility 
(iii) to apply, large flux variations are needed. For example, if the
X-ray emission of knot B is to fall on a direct extension of the
radio - optical spectrum with a monotonically increasing spectral index with
increasing frequency, its flux would have to be at least 5 times larger
than observed,
assuming no change in $\alpha_{\rm x}$ (Fig. 4). Harris, Biretta \& Junor
(1997, 1999) have found a secular decrease (significant at the 3$\sigma$ level)
in the intensity of knot A of some 16\% between 1992 and 1997 based on ROSAT
observations. Such small changes in X-ray flux are totally insufficient to
resolve the discrepancy between the X-ray spectra of knots A, B and D and
the extrapolations of their radio - optical spectra.
Further, the four separate determinations of the optical spectra of knots
D, A and B (Fig. 4) agree well, despite being taken over a period of
15 years (the observations of Biretta et al. [1991], Keel [1988],
P\'erez-Fournon et al. [1988] and Perlman et al. [2001] were taken in 1983,
1985, 1986 and 1998, respectively).
Thus the present data suggest that variations as large as needed do not occur.
However, given the limited angular 
resolution of the ROSAT observations, monitoring
with Chandra is highly desirable and should be able to resolve the issue,
especially if simultaneous optical spectra are obtained.

\section{Conclusions}

We have obtained high sensitivity, X-ray imaging - spectroscopy of the jet of
M87 with sub arc second angular resolution. Superficially, the X-ray jet
resembles that in the radio and optical bands, with all knots out to knot C
detected. However, there is a very strong trend for the ratio of X-ray to
either radio or optical flux to decline with increasing distance from the
nucleus. This strength of the near-nuclear jet in X-rays suggests that the X-ray
emission coincident with the nucleus may actually originate from the pc --
or sub-pc -- scale jet, rather than the accretion disk. There are clear 
morphological differences between the
radio/optical jet, on the one hand, and the X-ray jet, on the other. In
particular, some knots in the X-ray image are displaced from their
radio/optical counterparts towards the nucleus by tens of pc.

The spectra of the nucleus and all jet knots may be well described by
power-law spectra absorbed by cold matter. Only the nucleus shows clear
evidence for intrinsic absorption, with an equivalent hydrogen column
density of $\sim$ (3 -- 5) $\times$ 10$^{20}$ cm$^{-2}$. The knots in the
jet have photon indices, $\Gamma$, in the range 2.04 to 2.90, with average
$\Gamma$ $\simeq$ 2.4. The X-ray spectra of the jet knots are thus
considerably steeper than those at radio or optical wavelengths. Plotted
as $\nu$S$_{\nu}$, the spectra of the knots peak in or somewhat above
the optical - near infrared band. The X-ray photon index of the nucleus
is $\Gamma$ = 2.2$\pm$0.1, similar to the jet knots and again suggesting that
the nuclear X-ray emission originates from the jet.

We have discussed the process responsible for the X-ray emission. Synchrotron
self-Compton emission falls short by a factor of $\sim$ 100 -- 1,000 for an
equipartition field. The field needs to be $\sim$ 70 times below equipartition
to match the X-ray flux. Models
invoking inverse Compton scattering of the microwave background require values
of $\delta$ in the range 10 - 40 and an implausibly small angle
($\theta$ = 1$^{\circ}$ - 5$^{\circ}$) between the jet and the line of sight
(Harris \& Krawczynski 2002).
Furthermore, the spectral indices of the X-ray and radio emissions are
very different, in contrast with the similar spectral indices expected if the
synchrotron-emitting and inverse Compton scattering electron populations are
the same. For these reasons, we consider all inverse Compton models to
be extremely implausible.

The X-ray emission from the jet is almost certainly synchrotron radiation.
However, for at least three knots, the X-ray spectrum is not a simple
continuation of the radio -- optical spectrum. Instead, the spectrum must turn 
down at frequencies above the optical -- near infrared band, and then flatten
in X-rays.
The broad-band spectra of these knots are
remarkably similar to that of the western hot spot of Pictor A.
The optical, near infrared and X-ray data were not taken 
simultaneously, raising the possibility that our broad-band spectra may be
affected by variability. However, the magnitude of the knot variability
required to reconcile the observed spectra with simple synchrotron models,
in which the spectral index increases monotonically with increasing frequency,
is much larger than any yet observed. 
We briefly discuss the potential application to M87 of a model, originally
due to Burn (1973), in which the magnetic field is extremely inhomogeneous.
In this model, the X-ray emission originates from low energy electrons
temporarily in regions of very strong magnetic field; the X-ray spectrum is
related to the spectrum of the strength of the magnetic field rather than the
energy spectrum of the relativistic electons.
It is, however, very doubtful whether the large field filaments (with 
magnetic fields enhanced by a factor of $\sim$ 100) required can be created in
the jet plasma.
A third alternative is that
the X-ray synchrotron-emitting electrons/positrons are a separate
``population''
to those that emit the radio and 
optical synchrotron radiation (i.e. the energy spectrum and normalisation
of the X-ray emitting electrons at injection is not a simple continuation
of that of the radio- and optical-emitting electrons). 
If the last alternative is correct,
our X-ray spectra of the M87 jet provide additional support for the notion that
radio galaxies produce a hard ($\gamma$ = 2 - 2.5) spectrum of high energy
(E/m$_{\rm e}$c$^{2}$ $\sim$ 10$^{7-8}$) electrons and possibly positrons.

We are grateful to John Biretta
and Ismael P\'erez-Fournon for providing a 6 cm VLA map
and a V band image of the jet, respectively, and to
Carole Mundell for making a 2 cm map from the VLA archive. We thank the staff
of the Chandra Science Center, especially D. E. Harris and S. Virani, for their
help, C. D. Dermer for a valuable correspondence, E. S. Perlman for
helpful comments and D. E. Harris for pointing out an error in an earlier
version. This research was 
supported by NASA through grants NAG 81027 and NAG 81755.

\vfil\eject

\clearpage

\begin{deluxetable}{cccccc}
\tablecolumns{6}
\tablewidth{0pc}
\tablecaption{Distances of jet knots from the nucleus (arc secs)}
\tablehead{
\colhead{Waveband} & \colhead{D} & \colhead{E} & \colhead{F} & \colhead{A} &
\colhead{B}}
\startdata
X-ray & 2.6 & 5.7 - 5.9 & 8.1 &
12.2 - 12.3 & 14.1 \\
Optical & 3.1 & 6.1 & 8.4 - 8.5 &
12.2 - 12.3 & 14.2 \\
Radio & 3.0 - 3.1 & 5.8 - 6.0 & 8.6 - 8.8 & 12.3 - 12.4 & 14.3 - 14.5 \\
\enddata
\end{deluxetable}

\begin{deluxetable}{ccccccc}
\tablecolumns{7}
\tablewidth{0pc}
\tablecaption{Spectral fits to the X-ray emission of the
  nucleus and jet knots}

\tablehead{
\colhead{(1)} & \colhead{(2)} & \colhead{(3)} & \colhead{(4)} &
\colhead{(5)} & \colhead{(6)} & \colhead{(7)} \\ 
\colhead{Name} & \colhead{X\tablenotemark{a}, Y\tablenotemark{a}} &
\colhead{Region\tablenotemark{b}} & \colhead{N$_{\rm H}$\tablenotemark{c,}\,
\tablenotemark{d}} &
\colhead{$\Gamma$\tablenotemark{e}} & \colhead{K\tablenotemark{f}} &
\colhead{$\chi^2$} \\
\colhead{}     & \colhead{(arc sec)}                              &
\colhead{}                        & \colhead{($\times$ 10$^{20}$ cm$^{-2})$}  &
\colhead{}                          & \colhead{}                   &
\colhead{(degrees of freedom)}}

\startdata
Nuc (small)\tablenotemark{g} & 0, 0                               &
1$^{\prime\prime}$ sqr            & 6.1$^{+1.5}_{-1.4}$                       &
2.17$^{+0.10}_{-0.10}$              & 23.1$^{+1.8}_{-1.4}$        &
101.8 (106) \\
Nuc (large)\tablenotemark{g} & 0, 0                               &
2\farcs5 circ                     & 7.4$^{+1.5}_{-1.4}$                       &
2.25$^{+0.10}_{-0.09}$            & 40.5$^{+2.4}_{-2.9}$          &
145.2 (123) \\
D\tablenotemark{g}           & 2.6, 1.0                           &
1\farcs5 circ                     & 0.90$^{+2.0}_{-0.9}$          &
2.04$^{+0.16}_{-0.14}$            & 8.41$^{+0.72}_{-0.78}$        &
63.6 (60) \\
E                            & 5.6, 2.1                           &
1\farcs5 circ                     & 3.8$^{+3.4}_{-2.4}$           &
2.32$^{+0.26}_{-0.19}$            & 2.34$^{+0.50}_{-0.24}$           &
45.8 (39) \\
F                            & 7.7, 2.9                           &
1\farcs5 circ                     & 5.6$^{+2.4}_{-1.6}$           &
2.90$^{+0.44}_{-0.25}$            & 1.96$^{+0.31}_{-0.28}$           &
24.6 (32) \\
I                            & 10.1, 3.8                          &
1$^{\prime\prime}$ sqr            & 4.2$^{+7.0}_{-3.0}$           &
2.49$^{+0.50}_{-0.54}$            & 0.81$^{+0.24}_{-0.17}$           &
10.2 (13) \\
A\tablenotemark{g}           & 11.6, 4.4                          &
2\farcs0 circ                     & 1.3$^{+1.2}_{-1.1}$           &
2.32$^{+0.11}_{-0.10}$            & 20.9$^{+1.4}_{-1.1}$           &
152.4 (98) \\
B (small)                    & 13.3, 5.1                          &
1\farcs5 circ                     & 2.1$^{+3.0}_{-1.9}$           &
2.28$^{+0.24}_{-0.18}$            & 2.10$^{+0.38}_{-0.20}$           &
69.2 (39) \\
B (large)                    & 13.3, 5.1                          &
2\farcs5 circ                     & 1.7$^{+1.9}_{-1.6}$           &
2.24$^{+0.18}_{-0.09}$            & 3.97$^{+0.50}_{-0.29}$           &
72.0 (70) \\
C                            & 16.3, 6.5                          &
1\farcs6 $\times$ 1\farcs2 rect   & 3.3$^{+10.0}_{-3.3}$           &
2.54$^{+0.88}_{-0.51}$            & 0.64$^{+0.33}_{-0.10}$           &
14.3 (14) \\
C + G\tablenotemark{h}            & 17.3, 7.0                          &
3\farcs4 $\times$ 2\farcs0 rect   & 4.3$^{+4.3}_{-3.2}$           &
2.07$^{+0.30}_{-0.21}$            & 2.36$^{+0.70}_{-0.34}$           &
46.2 (46) \\

\tablenotetext{a}{Separation of center of region from the peak of the nuclear
source in R.A. (X) and Dec (Y).}
\tablenotetext{b}{Description of region from which spectrum was extracted.
The diameter is given for circles and the side lengths for squares and
rectangles.}
\tablenotetext{c}{Equivalent hydrogen absorbing column.}
\tablenotetext{d}{All errors are 90\% confidence for a single parameter of
interest.}
\tablenotetext{e}{Photon index.}
\tablenotetext{f}{Normalisation of power law model, in units of 10$^{-5}$
photons keV$^{-1}$ cm$^{-2}$ s$^{-1}$ at 1\,keV.}
\tablenotetext{g}{Spectra were obtained from 0\fs4 frame time observations.}
\tablenotetext{h}{This region includes only part of the knot referred to as G
in optical and radio observations.}

\enddata
\end{deluxetable}

\vfil\eject

\clearpage

\figcaption[f1.eps]
{Grey scale (upper panel) and contour (lower panel) representations of the X-ray
emission of the M87 jet in the 0.1 - 10 keV band. For the upper panel,
the image was rebinned to have a
pixel size of 0.1 arcsec and then smoothed by a Gaussian with a FWHM =
0\farcs5.
This rebinning was not done for the contour map to retain the original
number of counts pixel$^{-1}$.
The grey scale is
proportional to the square root of the X-ray brightness, and ranges from 0.02
times the peak
(white) to the peak (black). Contours are plotted at 20, 30, 60,
120, 180, 300, 400, 550, 700, 900 and 1000 counts pixel$^{-1}$. The field
plotted contains fainter, diffuse X-ray emission from M87 and the Virgo cluster;
this emission is not shown in order to emphasise the jet.
\label{Figure 1}}

\figcaption[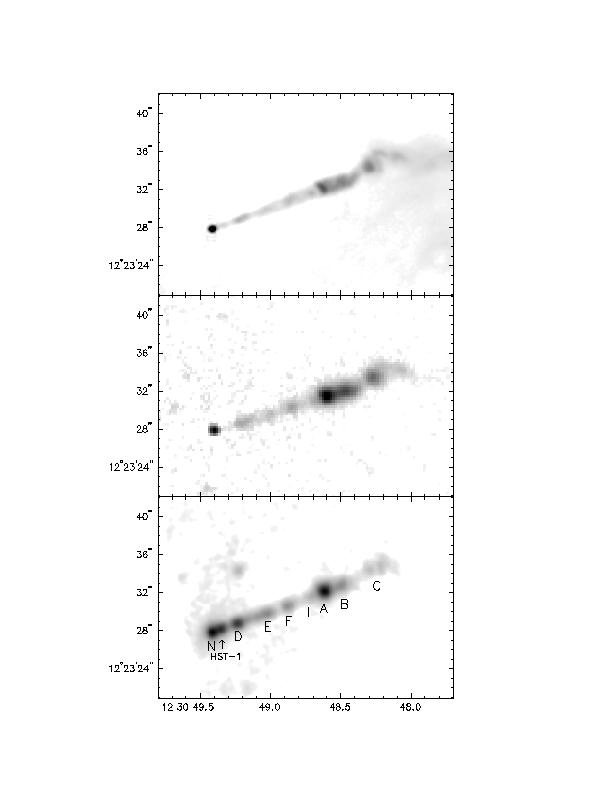]
{Grey scale representations of a 6 cm radio (top panel,
resolution 0$\farcs$4; Biretta, private communication), an optical V band 
(middle panel, resolution 0$\farcs$7; P\'erez-Fournon, private
communication) and the Chandra X-ray (bottom panel,
resolution 0$\farcs$7; 0.1 - 10 keV band)
image. In the
radio image, the grey scale is proportional to the square root of the brightness
and ranges from 0.0003 (white) to 0.03 (black)
Jy beam$^{-1}$. In the optical image,
the grey scale is also proportional to the square root of the brightness,
ranging from 20 to 2016 counts pixel$^{-1}$. The X-ray image is the same as in
Fig. 1. The labels in the lower panel refer to the knots vertically above the
label. N is the nucleus.
\label{Figure 2}}

\figcaption[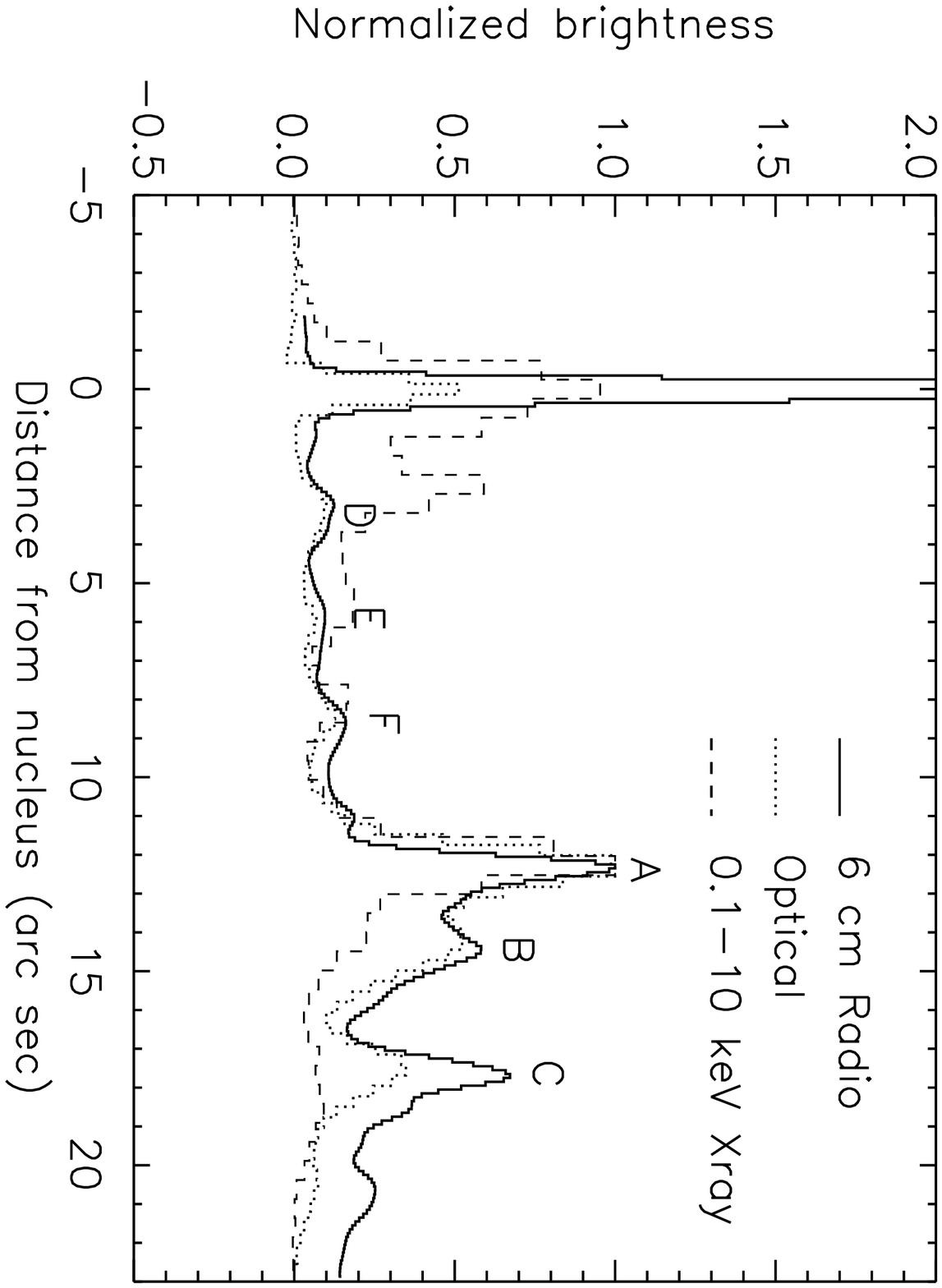]
{Plots of the brightness profiles along the M87 jet at 6 cm radio (solid line),
V band optical (dotted line) and 0.1 - 10 keV X-rays (dashed line).
Background emission from the surrounding diffuse X-ray emission has been
subtracted from the X-ray profile. In producing these profiles, emission over a
region 3$^{\prime\prime}$ perpendicular to the jet axis was summed at each
location at each 
wavelength. It is noteable that the X-ray knots D and F, and possibly knot E,
are displaced towards the nucleus from their radio and optical counterparts.
\label{Figure 3}}

\figcaption[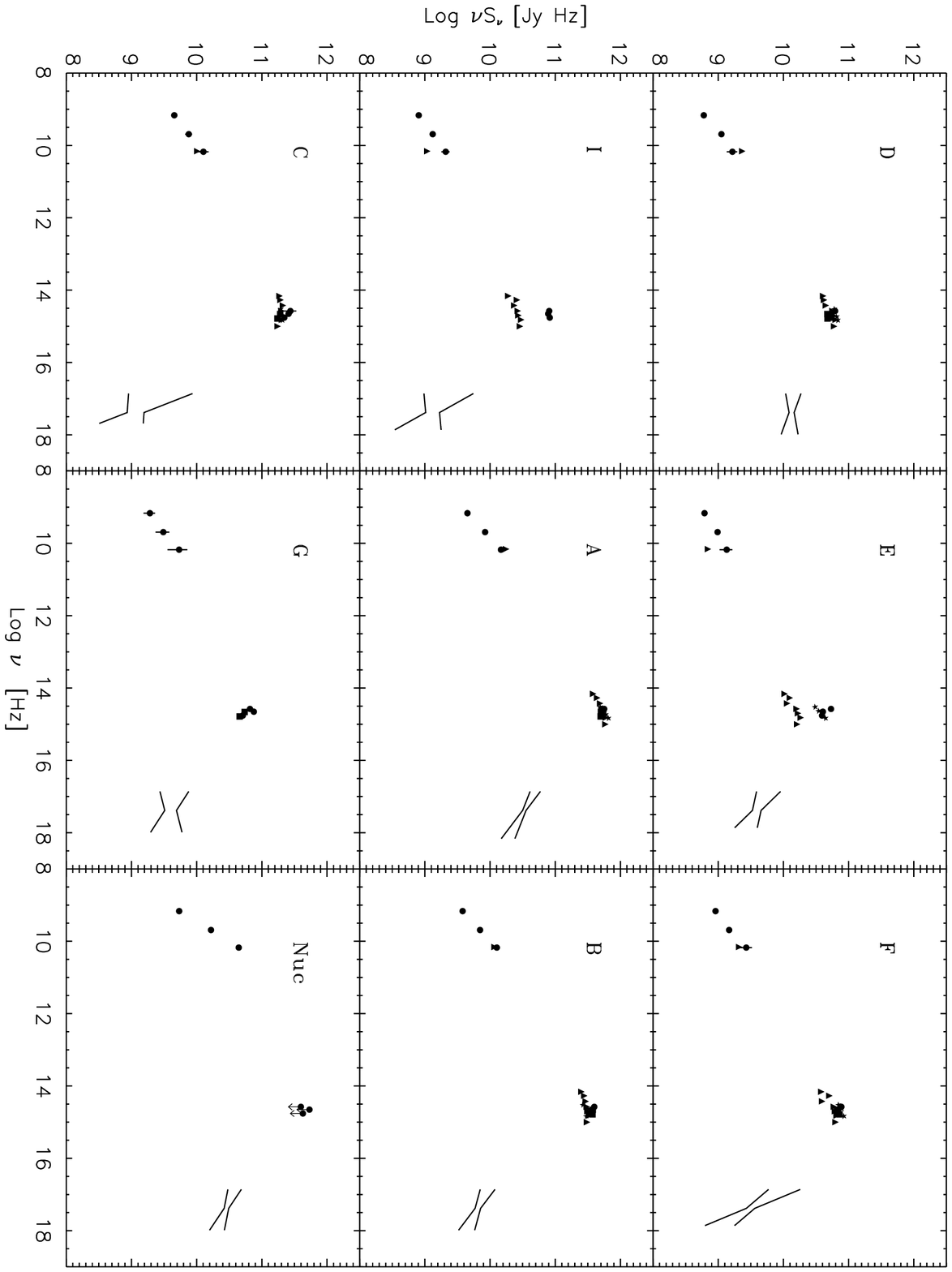]
{The radio through X-ray spectra of the various knots in the jet and the
nucleus of M87. Filled squares are from P\'erez-Fournon et al. (1988),
asterisks are from Keel (1988), filled circles are from Biretta, Stern \&
Harris (1991) and filled triangles are from Perlman et al. (2001). The
``bow tie'' represents the X-ray spectrum, with its 90\% confidence
uncertainties, from our Chandra observations. The differences in the measured
optical spectra of knots E and I may result from overestimation of the flux of
these
weak knots in ground-based observations (asterisks and circles). The X-ray
spectrum given for knot G contains flux from the optical/radio locations of
both knots C and G and does not represent the X-ray spectrum of knot G alone.
\label{Figure 4}}

\figcaption[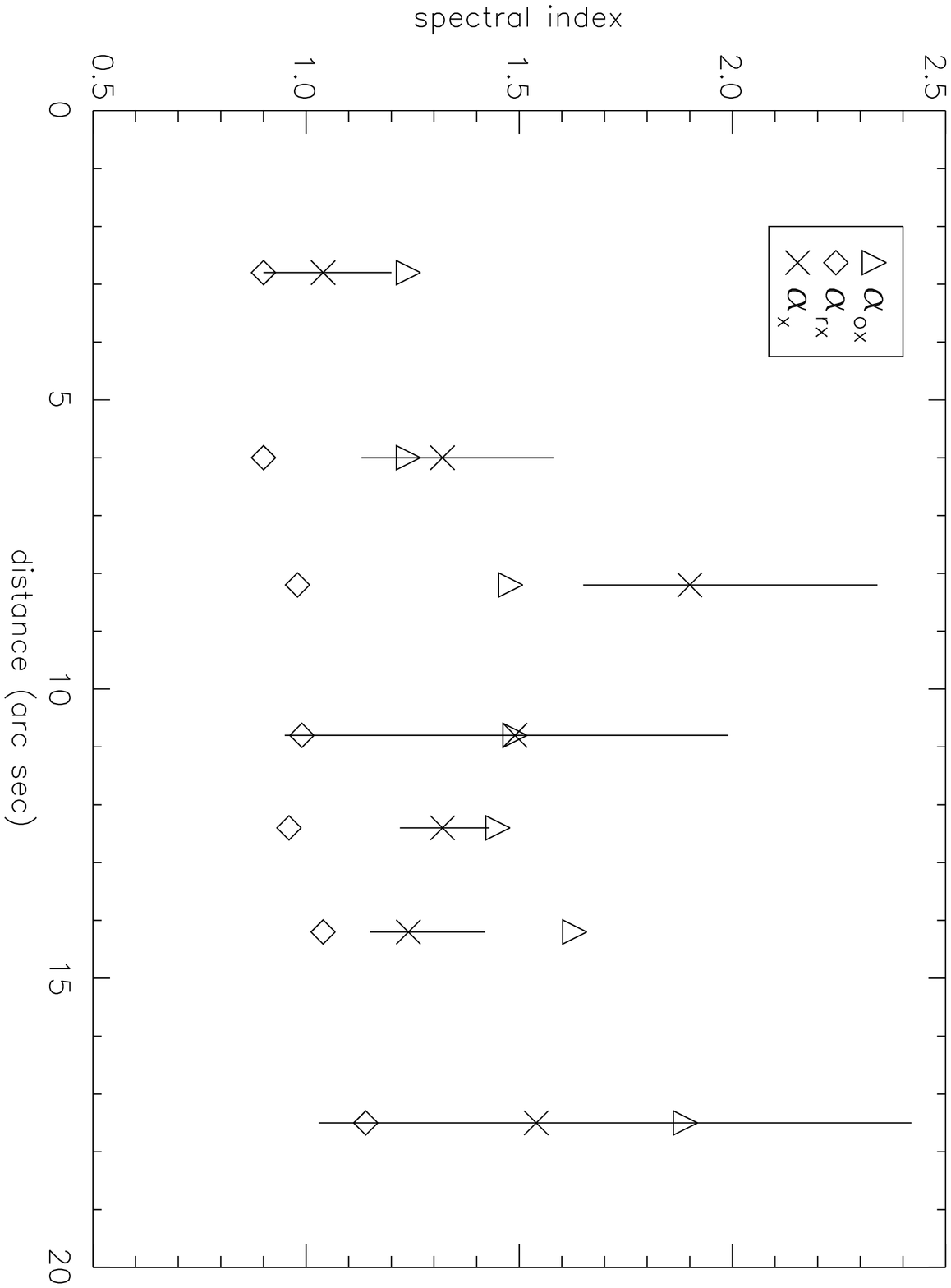]
{Plots of the optical to X-ray ($\alpha_{\rm ox}$), radio to X-ray
($\alpha_{\rm rx}$) and X-ray ($\alpha_{\rm x}$) spectral indices for the knots
in the jet as a function of distance from the nucleus. The key is given in the
upper left. Here optical means 6000 \AA, radio 2 cm and X-ray 1 keV. The
errors on $\alpha_{\rm ox}$ and $\alpha_{\rm rx}$ are smaller than the symbol
sizes and are not shown.
The error bars represent the 90\% confidence errors on $\alpha_{\rm x}$.
\label{Figure 5}}

\clearpage


\begin{references}{}

\reference{} Band, D. L. \& Grindlay, J. E. 1985, ApJ, 298, 128

\reference{} Band, D. L. \& Grindlay, J. E. 1986, ApJ, 308, 576

\reference{} Begelman, M. C., Blandford, R. D. \& Rees, M. J. 1984, Rev. Mod.
Phys. 56, 255

\reference{} Biretta, J. A. 1999, In The Radio
Galaxy M87, Eds, H.-J. R\"oser and K. Meisenheimer, p 159 (Springer: Berlin)

\reference{} Biretta, J. A., Stern, C. P. \& Harris, D. E. 1991, AJ, 101, 1632

\reference{} Biretta, J. A., Sparks, W. B. \& Macchetto, F. D. 1999, ApJ, 520,
621

\reference{} B\"ohringer, H. et al. 2001, A\&A, 365, L181

\reference{} Burbidge, G. R., Jones, T. W. \& O'Dell, S. L. 1974, ApJ, 193, 43

\reference{} Burn, B. J. 1973, MNRAS, 165, 421

\reference{} Cavallo, G., Horstman, H. M. \& Muracchini, A. 1980, A\&A, 86, 36

\reference{} Dermer, C. D. 1995, ApJ, 446, L63

\reference{} Dermer, C. D., Sturner, S. J. \& Schlickeiser, R. 1997, ApJS, 109,
103

\reference{} Ginzburg, V. L. 1957, Uspekhi Fiz. Nauk, 62, 37

\reference{} Harris, D. E., Biretta, J. A. \& Junor, W. 1997, MNRAS, 284, L21

\reference{} Harris, D. E., Biretta, J. A. \& Junor, W. 1999, In The Radio
Galaxy M87, Eds, H.-J. R\"oser and K. Meisenheimer, p 319 (Springer: Berlin)

\reference{} Harris, D. E. \& Krawczynski, H. 2002, ApJ (in press)

\reference{} Jaffe, W. J. \& Perola, G. C. 1973, A\&A, 26, 421

\reference{} Junor, W. \& Biretta, J. A. 1995, AJ, 109, 500

\reference{} Kardashev, N. S. 1962, Sov. Astron - AJ, 6, 317

\reference{} Keel, W. C. 1988, ApJ, 329, 532

\reference{} Kirk, J. G. \& Dendy, R. O. 2001, Journal of Physics G27, 1589

\reference{} Kirk, J. G. 2002, In Particles and Fields in Radio Galaxies,
ASP Conference Series, Eds. R. A. Laing \& K. M. Blundell (Astronomical 
Society of the Pacific) (in press)

\reference{} Macchetto, F. D., Marconi, A., Axon, D. J., Capetti, A.,
Sparks, W. \& Crane, P. 1997, ApJ, 489, 579

\reference{} Mannheim, K., Kr\"ulls, W. M. \& Biermann, P. L. 1991, A\&A,
251, 723

\reference{} Marscher, A. P., Marshall, F. E., Mushotzky, R. F., Dent, W. A.,
Balonek, T. J. \& Hartman, M. F. 1979, ApJ, 233, 498

\reference{} Marscher, A. P. 1983, ApJ, 264, 296

\reference{} Meisenheimer, K., R\"oser, H.-J. \& Schl\"otelburg, M. 1996,
A\&A, 307, 61

\reference{} Neumann, M., Meisenheimer, K., R\"oser, H.-J. \& Fink, H. H.
1997, A\& A, 318, 383

\reference{} Owen, F. N., Hardee, P. E., \& Cornwell, T. J. 1989, ApJ, 340,
698

\reference{} P\'erez-Fournon, I., Colina, L., Gonz\'alez-Serrano, J. I. \&
Biermann, P. L. 1988, ApJ, 329, L81

\reference{} Perlman, E. S. Biretta, J. A., Sparks, W. B., Macchetto, F. D.
\& Leahy, J. P. 2001, ApJ, 551, 206

\reference{} Schreier, E. J., Gorenstein, P. \& Feigelson, E. D. 1982,
ApJ, 261, 42

\reference{} Stark, A. A., Gammie, C. F., Wilson, R. W., Bally, J., Linke, R. 
A., Heiles, C. \& Hurwitz, M. 1992, ApJS, 79, 77 

\reference{} Tavecchio, F., Maraschi, L., Sambruna, R. M. \& Urry, C. M. 2000,
ApJ, 544, L23

\reference{} Tonry, J. L. 1991, ApJ, 373, L1

\reference{} Tregillis, I. L., Jones, T. W. \& Ryu, D. 2001, ApJ, 557, 475

\reference{} Whitmore, B. C., Sparks, W. B., Lucas, R. A., Macchetto, F. D. \&
Biretta, J. A. 1995, ApJ, 454, L73

\reference{} Wilson, A. S., Young, A. J. \& Shopbell, P. L. 2001, ApJ, 547, 740

\reference{}

\reference{}

\end{references}
\end{document}